# Prototype of Pulse Digitizer and Readout Electronics for CDEX-10 in CJPL

**Jinfu Zhu** [a,b], **Hongming Li** [c], **Tao Xue** [a,b, *], **Liangjun Wei** [d], **Jianmin Li** [a,b]

[a] *Key Laboratory of Particle & Radiation Imaging (Tsinghua University), Ministry of Education,*
*Beijing, China*
[b] *Department of Engineering Physics, Tsinghua University,*
*Beijing, China*
[c] *Tsinghua National Laboratory for Information Science and Technology (TNList),*
*Beijing, China*
[d] *NUCTECH Company Limited,*
*Beijing, China*

*E-mail:* xuetaothu@tsinghua.edu.cn

ABSTRACT: The CDEX (China Dark matter Experiment) is currently in the stage to upgrade to approximately 10 kg HPGe (High Purity Germanium) detectors (CDEX-10) in CJPL (China Jinping Underground Laboratory) and aims to detect the WIMP (Weakly Interacting Massive Particles). Six analog signals and one inhibited signal are outputted from preamplifier by one HPGe detector (~1 kg) of CDEX-10. The analog signals are conditioned with two 6-μs and one 12-μs shaping amplifiers, and three timing amplifiers respectively. The inhibited signal is used for trigger. The analog signals from shaping and timing amplifiers are digitized to obtain energy spectrum and rise-time distribution. The rise-time distribution is used to discriminate surface, bulk, and very bulk events, which can help to supress the background level in dark matter detection. Composed of commercial modules, the previous DAQ (Data Acquisition Systems) is insufficient in discriminating between bulk and very bulk events due to the limited sampling rate (100 MSPS). Besides, as number of channel increases, the system will be simplified by integrating the trigger logic in FPGA (Field Programmable Gate Array). In this paper, a prototype named RAIN4HPGe is designed with pulse digitizer and readout electronics, integrating 14-Bit 100 MSPS ADCs (Analog-to-Digital Converters), 12-Bit 1 GSPS ADCs, FPGA etc.. The RAIN1000Z1 readout module based on ZYNQ SoC (System on Chip) is used for readout over Gigabit Ethernet. Better energy resolution, e.g. improving from 558 eV to 474 eV @122 keV, can be achieved by RAIN4HPGe compared to previous DAQ (Data Acquisition Systems) with commercial solution. Moreover, the discrimination performance of bulk, and very bulk events has been improved by RAIN4HPGe with 1 GSPS ADC.

---

\* Corresponding author.

# Contents

# 1. Introduction

The CDEX [1] (China Dark matter EXperiment) in CJPL (China Jinping Underground Laboratory) [2] is located in Xichang, Sichuan province, China. CDEX-1 deploys one kg-scale mass pPCGe (p-type Point Contact Germanium) detector enclosed by a NaI(Tl) crystal scintillator as anti-Compton detector [3]. The DAQ (Data Acquisition System) mainly consists of NIMs (Nuclear Instrument Modules) and VME (Versa Module Eurocard) systems, e.g. V1724 (8-channel 14-Bit 100 MSPS digitizer) [4], N93B (dual timer) [5], and N840 (8-channel leading edge discriminator) [6], etc. The new stage of CDEX has increased HPGe detectors to 9 modules (with a total detector mass of ~10 kg) and will scale up to ~100 kg in the future [7] [8]. Due to the slight difference of rise-time between bulk and very bulk events, it is insufficient for pulse shape discrimination by the previous DAQ with limited sampling speed (100 MSPS). Very bulk events occur near the p+ point surface of pPCGe, which are usually from radioactive materials of electronic devices. The improvement of discrimination of bulk and very bulk events helps to further suppress the background level [9].

The homemade RAIN4HPGe is designed with high sampling speed pulse digitizer (8-channel 14-Bit 100 MSPS ADCs (Analog-to-Digital Converters) and 4-channel 12-Bit 1 GSPS ADCs), high throughput buffer and high bandwidth readout electronics. RAIN4HPGe can realize pulse digitalization, trigger and readout from detector arrays, and can be plugged into a 6U-form-factor crate, transmit data to PC through Gigabit Ethernet interface. It is also more flexible to update the hardware, firmware, and software online.

This paper is organized as follows. Section 2 describes the DAQ system architecture of CDEX-10 with the previous DAQ and RAIN4HPGe, respectively, and then introduces hardware, software and firmware design of RAIN4HPGe. Section 3 presents the performance test of high throughput, high bandwidth external DDR3 SDRAM (Double-Data-Rate Synchronous Dynamic

Random Access Memory generation 3), gamma-ray sources experiments, and comparison of performance between the previous DAQ and RAIN4HPGe. Conclusions and future work will be given in Section 4.

## 2. Design and implementation

### 2.1 System architecture of pulse digitizer and readout electronics in CDEX-10

Figure 1 and Figure 2 depict the overall system architecture of CDEX-10 experiment with previous DAQ and RAIN4HPGe respectively. CDEX-10 has deployed three triple-element pPCGe detector strings (C10A, B, C) directly immersed in liquid nitrogen (LN$_2$) [8]. Each pPCGe detector works independently. For one HPGe detector, e.g. C10B Ge1 (~1 kg), there are six identical analogy signals from the JFET (Junction gate Field-Effect Transistor) reset type preamplifiers and one inhibited signal. Waveforms with 6-µs shaping time are used to calibrate energy and those with ~300-ns shaping time are for rise-time measurement. The inhibited signal and the output of the JFET preamplifier are shown in Figure 3, note the step signal with a long time grow slope offset is one hit and the abrupt return to -2V is the reset signal. The reset period of the JFET preamplifier is ~1.2 s.

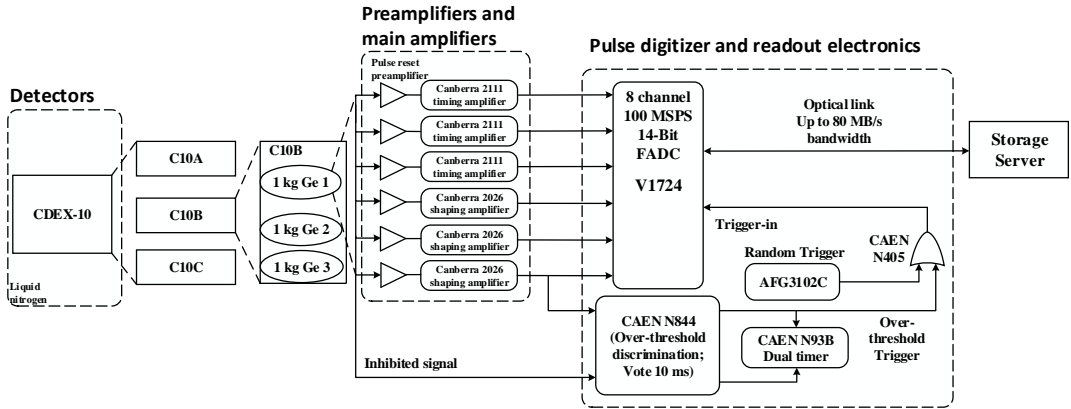

**Figure 1.** The DAQ system architecture of the CDEX-10 experiment with the previous DAQ.

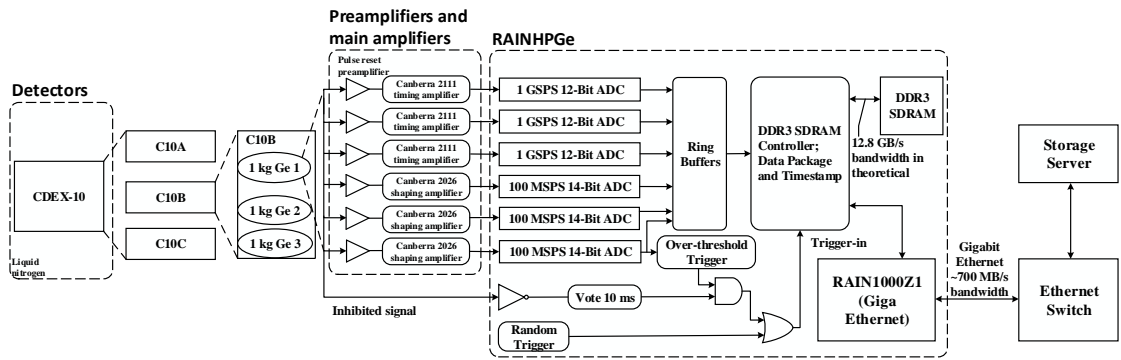

**Figure 2.** The DAQ system architecture of the CDEX-10 experiment with the homemade RAIN4HPGe prototype.

The inhibited signal is a positive pulse (~0.9 ms width) synchronized with reset of the preamplifier. Three signals from the JFET preamplifiers are distributed to shaping amplifiers (Canberra 2026) with high gain in low-energy region (0-12 keV, shaping time 6-μs and 12-μs) and low gain in high-energy region (0-240 keV, shaping time 6-μs), respectively. The other three signals are loaded to timing amplifiers (Canberra 2111, shaping time ~300 ns), two with high gain in the medium-energy region (0-20 keV) and one with low gain in high-energy region (0-240 keV).

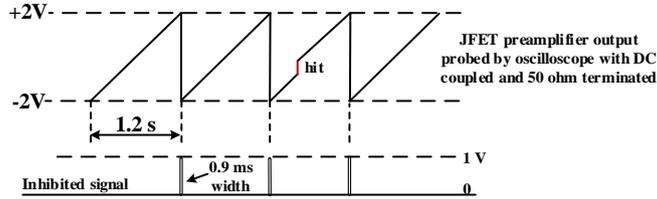

**Figure 3.** The JFET reset type preamplifier's output (with one hit) and inhibited signal.

In Figure 1, readout and triggering electronics are composed of commercial digitizer (V1724) and other NIMs or VME systems, analogy signals from shaping and timing amplifiers are digitized by 14-Bit 100 MSPS ADCs. The CAEN N844 is used to discriminate the pulse from 6-μs shaping amplifier (0-12 keV energy range) and votes 10 ms, i.e. force to make the over-threshold trigger output logic '0', hold for 10 ms after inhibited signal and outputs a ~1V pulse. Periodic (0.05 Hz) signals from arbitrary waveform generator (Tektronix AFG3102C) is used as random trigger to monitor noise level of the total system, as it's random relative to the moment of the physical event. The CAEN N93B records the moments of over-threshold trigger and random trigger. The CAEN N405 is used as OR gate. When the Trigger-in is logic '1', the six pulses (~120-μs length) are digitized and buffered, then transmitted to storage server via optical link.

In Figure 2, RAIN4HPGe, which combines the 100 MSPS 14-Bit ADCs and 1 GSPS 12-Bit ADCs, is designed for pulse digitalization of two kinds of signals, i.e., "slow" pulses from shaping amplifiers and "fast" from timing amplifiers. Waveform data from ADCs is buffered in the *Ring Buffers* in Kintex-7 FPGA (Field Programmable Gate Array), and then saved to high throughput external DDR3 SDRAM controlled by the FPGA. The trigger is implemented real-time in the FPGA based on inhibited signal, the over-threshold trigger, and the 0.05 Hz "random" trigger. After the NOT gate, inhibited signal will be voted by 10 ms, i.e. keep logic '0' for 10 ms to the AND gate after inhibited pulse generates. When the *Trigger-in* is logic '1', data is read from DDR3 SDRAM, stamped with timer maintained in FPGA and transmitted to PC with RAIN1000Z1 readout module [10] [11] based on ZYNQ SoC (System on Chip) through Gigabit Ethernet interface. Commercial Ethernet switches and servers are used for online transmission and storage.

## 2.2 Hardware design and challenges

Kintex XC7K325T FPGA is interfaced to ADCs via LVDS (Low-Voltage Differential Signal) for high speed data transmission and via SPI (Serial Peripheral Interface) for slow control. A 64-Bit 1600 MHz DDR3 SDRAM is integrated in the system and associated with the FPGA for high-speed data throughput and large capacity data buffer.

RAIN1000Z1 module based on ZYNQ SoC has been developed for compact readout applications. It is used for high speed data exchange to PC or data servers. Embedded Linux runs in the readout module and associated software based on TCP/IP is developed for data transmission with Gigabit Ethernet interface.

One of design challenges is high-speed PCB (Printed Circuit Board) layout for 1 GSPS ADC and 1600 MHz 64-Bit DDR3 SDRAM. As calculated, overall input data bandwidth is up to 59.2 Gbps, which includes the data from 8-channel 14-Bit 100 MSPS ADC and 4-channel 12-Bit 1 GSPS ADC. High bandwidth waveform data from external ADCs will be saved in the internal *Ring Buffers* firstly and waits for trigger signal to select according to physical requirements, a large capacity high data throughput external buffer is also critical for requirement of correlated events with long time continuous no-interrupt digitalization and buffer.

Length match and impedance control of high-speed signal are critical for hardware design. The fly-by topology is used for the address and control signals of DDR3 SDRAMs, and point-to-point topology is used for data signals. Figure 4 shows the prototype's picture of the 6U pulse digitalization and readout board.

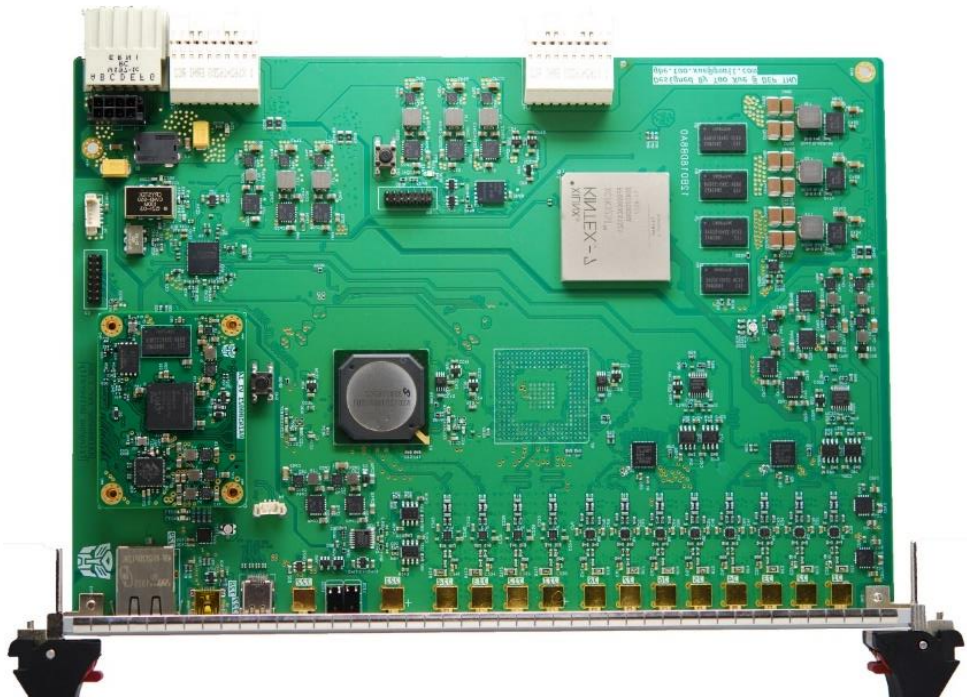

**Figure 4.** Picture of the readout electronics' 6U-form-factor prototype board, RAIN4HPGe.

For high-speed hardware PCB layout designs, SI (signal integrity) is the key issue. With fly-by topology, high-speed 1600 MHz memory interfaces for DDR3 SDRAM should maintain the signal traces on the PCB as 40-Ohm impedance-controlled and length-matched, and simultaneously the flight time compensated for in the silicon package.

ADC driver ADA4927 and 14-Bit 100 MSPS AD9253 (quad channels, theoretically 0.4 W power consumption) from Analog Devices are used for pulse digitalization of slow signals and PMT's signals. ADC driver LMH6554 driver and 12-Bit 1 GSPS ADC12D1000 (dual channels, 3.6 W theoretical power consumption) from TI are used for pulse digitalization of fast signals. Precise DAC AD5686 from Analog Devices is used for input bias offset tuning.

Power supply for high speed ADCs and analog components are critical in the RAIN4HPGe. Considering the requirements of high power efficiency and low noise, strategy of switched

DC/DC regulator accompanied by ultra-low noise LDO (Low Drop Output) linear voltage regulator is used for analog power supply. TPS62130 (about 90% efficiency at 3.0A) and TPS74401 (only 115 mV drop at 3.0A) or TPS7A7300 (240 mV drop at 3.0A) from TI are used. For example, 1.92V is generated from 5V with the TPS62130 and a 1.8V analog power rail is generated from 1.92V with TPS74401. At 3.0A, around an 84% typical power efficiency and 30 $\mu V_{RMS}$ noise can be achieved for analog power rail. For negative power rail, TPS62130 is used in negative output mode, and TPS7A4701 with TPS7A3301 are used for combined positive and negative linear voltage regulators.

System clock is generated by LMK04806 with an external clock distributed from rubidium clock or on-board CCHD-575 ultra-low phase noise oscillator from Crystek Corp. Low phase noise VCXO CVHD-950 voltage-controlled oscillator is associated with LMK04806 for jitter cleaner. Integrated synthesizer and VCO ADF4360-7 generate a 1 GHz clock for ADC12D1000.

### 2.3 Software and firmware design

The Vivado IDE (Integrated Development Environment) from Xilinx is used for system design and firmware design.

For RAIN1000Z1 readout module, Ubuntu 14.04.3 LTS with the cross compiler ARM GCC (GNU Compiler Collection) is used for program development based on embedded Linux. A user interface program, which is developed based on GCC in Ubuntu, is used for real-time data storage, monitor and analysis. Some dedicated device drivers [12] [13] are developed for I/O control, including the control of DC offset tuning and ADC parameters. EPICS IOC is implemented and used for distributed slow control [14].

Firmware is designed with Verilog HDL (Hardware Description Language). Ring buffers and associated control function for trigger are also implemented within the FPGA. A memory controller IP core with DDR3 SDRAM interface supported [15] is implemented within the firmware in order to interface with external 1600 MHz 64-Bit DDR3 SDRAM, which is used as external buffers.

For off-line data storage and performance analysis, a LabVIEW-based user interface is developed to analyse the ENOB performance on PC or server.

## 3. Performance test and gamma-ray sources experiments

### 3.1 External buffer performance test

Four 16-Bit 1600MHz DDR3 SDRAMs, i.e., K4B2G1646C from SAMSUNG, are used as external buffers [16]. Every DDR3 SDRAM chip has capacity of 256 MB and the overall external buffers' storage is up to 1 GB. Theoretical bandwidth of the external buffers is calculated as eq. (3.1).

$$Theoretical\ bandwidth(Gbps) = freq(GHz) \times width(bit) = 1.6 \times 4 \times 16 = 102.4 . \qquad (3.1)$$

DDR controller in the firmware needs to generate periodic fresh signals, as the data stream between FPGA and DDR3 SDRAM is not sustainable in the physical interface. It is necessary to know the actual bandwidth of external buffers, i.e., the read and write efficiency of DDR3 SDRAM [17]. In our test, auxiliary program logic was designed with the DDR controller. IP core of the DDR controller provides signals of *user clock*, *app_wdf_rdy*, *app_rdy_out*, etc., in the user interface, which helps to check whether the data is valid or not. Several parameters are defined as follows.

No. of test write pattern: $N_w$

No. of cost user clock cycle to write: $N_{cw}$

No. of test read pattern: $N_r$

No. of cost user clock cycle to read: $N_{cr}$

The read and write efficiency of DDR3 SDRAM is defined by eq. (3.2) and calculated by eq. (3.3). A sequence of numbers (0, 1, 2, 3, ... 268435455) is used as test patterns, which are written into or read from their corresponding physical addresses. Therefore, the total number of test patterns is 268435456 (0x10000000 in hexadecimal).

$$\eta_{\text{write efficiency}} = \lim_{N_w \to \infty} \frac{N_w}{N_{cw}} \times 100\% \ , \tag{3.2}$$

$$\eta_{\text{read efficiency}} = \lim_{N_r \to \infty} \frac{N_r}{N_{cr}} \times 100\% \ .$$

$$\eta_{\text{write efficiency}} \approx \frac{268435456}{316981745} \times 100\% = 84.68\% \ , \tag{3.3}$$

$$\eta_{\text{read efficiency}} \approx \frac{268435456}{292196993} \times 100\% = 91.87\% \ .$$

According to eq. (3.4), actual bandwidth of external buffers can be estimated to be 87 Gbps, approximately, which can meet the bandwidth requirements for ADCs' input.

$$Actual\ bandwidth = Theoretical\ bandwidth \times Efficiency \tag{3.4}$$

The lengths of waveforms digitized by 14-Bit 100 MSPS ADC and 12-Bit 1 GSPS ADC are 120 μs and 16 μs respectively. The maximum events stored by DDR3 SDRAM is about ~7585 according to eq. (3.5). Background event rate of CDEX-10 is estimated ~34 cpkkd (count per kg per keVee per day), and the physical requirement for DAQ is 1000 cpkkd.

$$The\ maximum\ events = \frac{Storage\ of\ DDR3\ SDRAM}{\sum Data\ Bandwidth\ of\ ADCs \times Record\ Length} = \frac{8Gb}{3 \times 100Msps \times 14b \times 120\mu s + 3 \times 1Gsps \times 12b \times 16\mu s} \approx 7585 \tag{3.5}$$

## 3.2 Typical waveforms and energy spectrum

The typical 6-μs shaping waveform and its power spectrum are shown in Figure 5. The waveform is from shaping amplifier (Canberra 2026) and digitized by 14-Bit 100 MSPS ADC. The typical "fast" waveform and its power spectrum from timing amplifier (Canberra 2111), digitized by 12-Bit 1 GSPS ADC, are depicted in Figure 6.

Different gamma-ray sources ($^{241}$Am, $^{109}$Cd, and $^{57}$Co) are used to calibrate energy spectrum (the height of 6-μs shaping pulse is used) and calculate energy resolution. As portrayed in Figure 7, the energy resolution is (474 ± 8) eV @ 122 keV. The energy spectrum is fitted by the one-degree polynomial and the correlation coefficient $R^2$ is better than 0.999999.

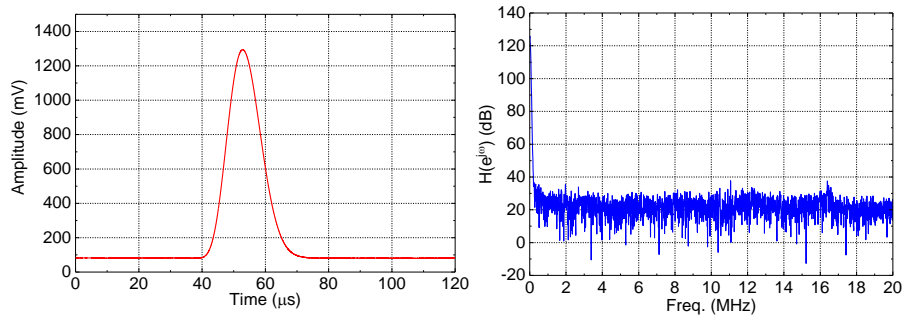

**Figure 5.** The waveform from the shaping amplifier and its power spectrum.

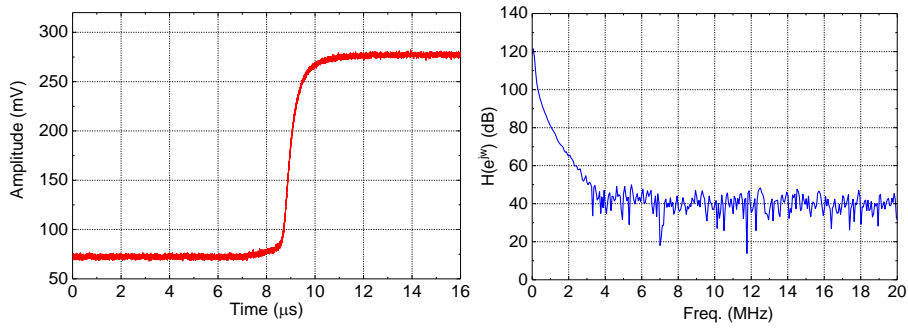

**Figure 6.** The waveform from the timing amplifier and its power spectrum.

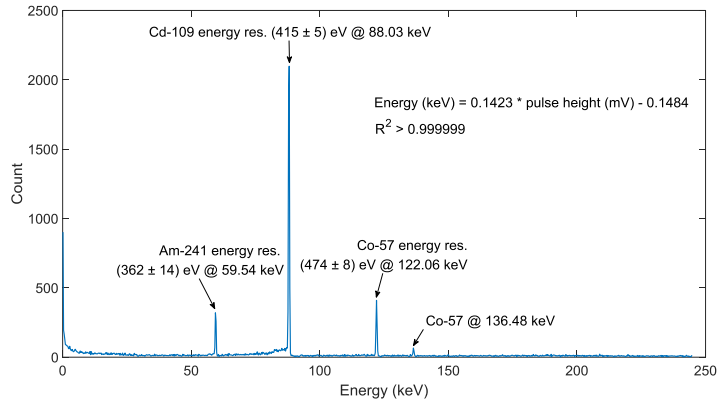

**Figure 7.** The gamma-ray energy spectrum, the height of 6-µs shaping pulse is used.

### 3.3 Discrimination of surface, bulk, and very bulk events

Discrimination based on rise-time distribution is critical in dark matter detection. The rise-time is defined as the time interval from 10% to 90% amplitude. The rise-time difference is due to the various location of charge collection in 1 kg-scale mass pPCGe.

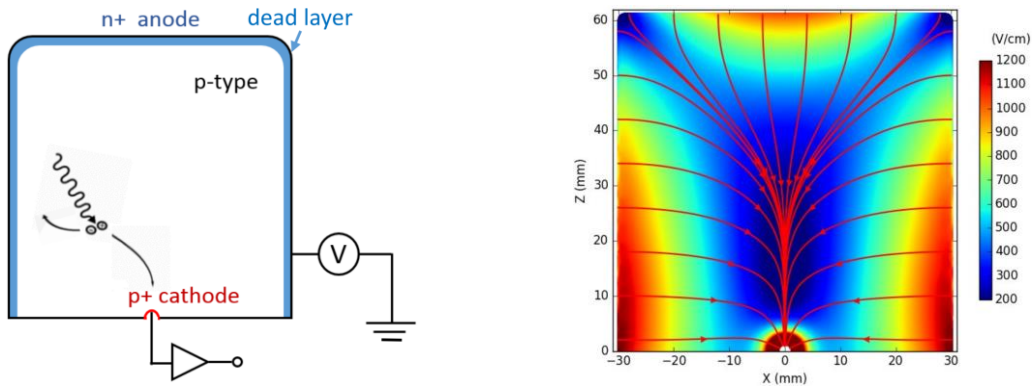

**Figure 8.** (a) The structure of 1 kg pPCGe detector.       (b)    Electric field simulation.

As shown in Figure 8 (a), there is a dead layer (in colour blue) at the surface fabricated by lithium diffusion. Events that occur at the surface (called surface events) have a slower drift time and incomplete charge collection, comparing with those occur in the bulk (called bulk events). Figure 8 (b) depicts the electric field simulated result of pPCGe detector. The arrows in lines are motion directions of carriers. The charge collection of carriers is driven by electric field. There exists a weak electric area (in colour blue) in the centre of detector. If interactions occur near the p-point face, the charges will be collected directly without going through the weak electric area, leading to a faster drift time. This is the main origin of very bulk events.

Rise-time information of waveforms is retained by timing amplifiers. The distributions of rise-time as energy are shown in Figure 9 (a) and (b), which are digitized by RAIN4HPGe and V1724 respectively. It presents that surface, bulk, and very bulk events can be discriminated sufficiently above ~20 keV by RAIN4HPGe, especially for bulk and very bulk events. Waveforms digitized by RAIN4HPGe have finer discrimination than that by V1724, owing to its faster sampling rate (1 GSPS). The preliminary Gaussian fittings of rise-time distributions are processed and their FWHMs (Full Width at Half Maximum) are compared in Figure 10.

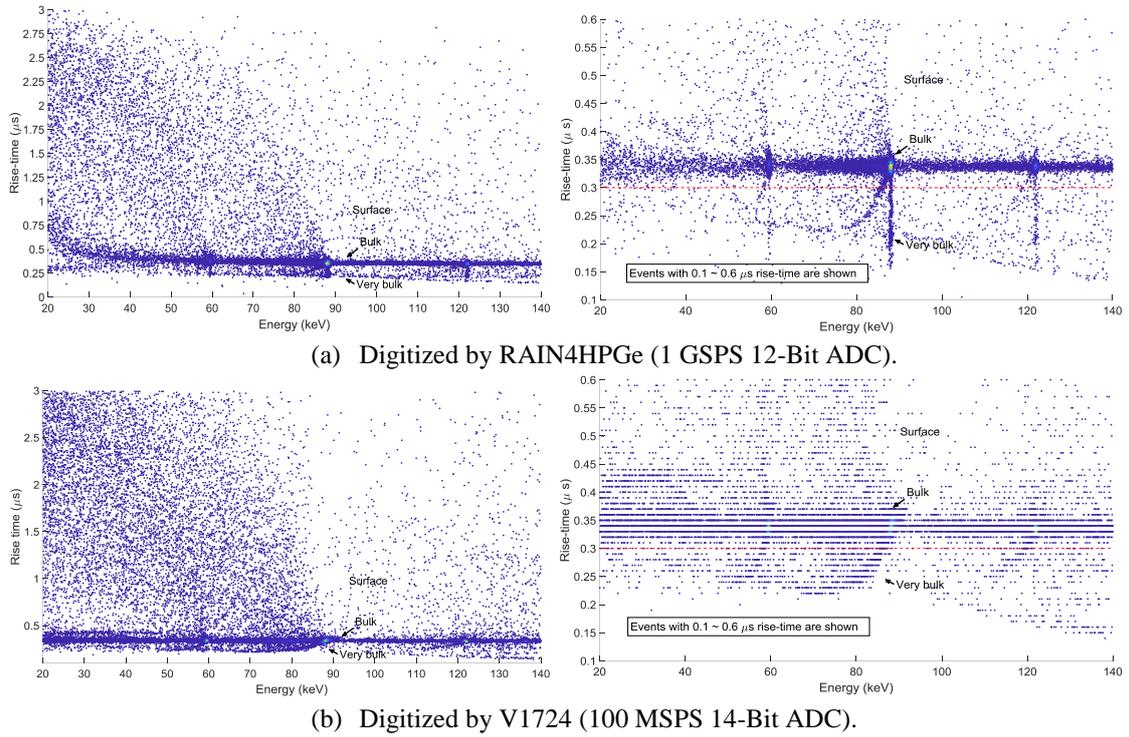

(a) Digitized by RAIN4HPGe (1 GSPS 12-Bit ADC).

(b) Digitized by V1724 (100 MSPS 14-Bit ADC).

**Figure 9.** The discrimination of bulk, very bulk, and surface events.

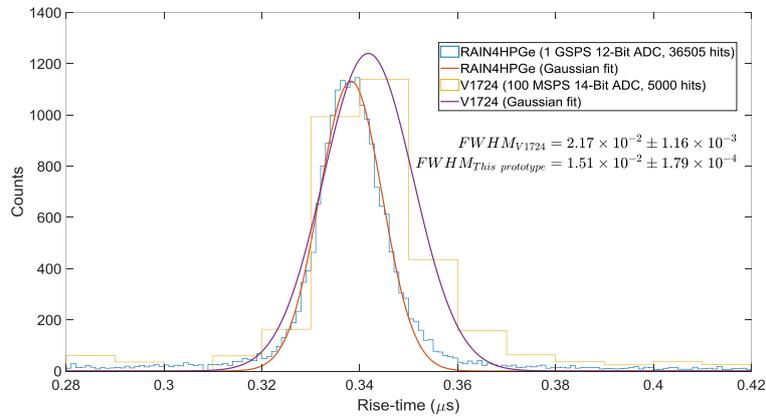

$$FWHM_{V1724} = 2.17 \times 10^{-2} \pm 1.16 \times 10^{-3}$$
$$FWHM_{This\ prototype} = 1.51 \times 10^{-2} \pm 1.79 \times 10^{-4}$$

**Figure 10.** The rise-time distributions and their Gaussian fit.

Typical surface, bulk, and very bulk waveforms digitized by RAIN4HPGe and their power spectrum are drawn in Figure 11. As described, surface waveforms have longer rise-time than bulk and very bulk waveforms. The difference of bulk and very bulk waveforms is mainly at the first corner of leading edge due to the shorter drift time of very bulk events. The maximum bandwidth of bulk and very bulk events' typical waveforms is ~8 MHz and the surface events has narrow frequency bandwidth ~2 MHz due to its slower rising rate.

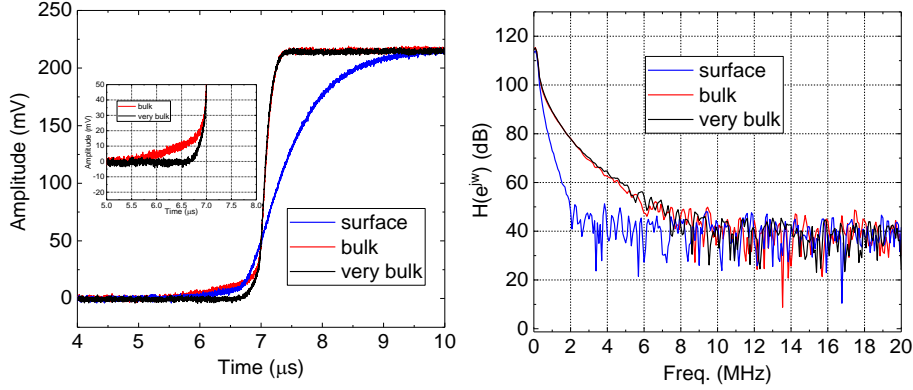

**Figure 11.** The typical waveforms of surface, bulk, and very bulk and their power spectra.

### 3.4 Performance comparison of the previous DAQ and RAIN4HPGe

The ADC ENOB (effective number of bits), readout bandwidth, and experimental results etc. of RAIN4HPGe and previous DAQ are listed in Table 1. For RAIN4HPGe, the full-scale of analog input is decided by ADC driver and analog input range of ADC. The bandwidth of analogy input depends on anti-aliasing filter after ADC driver. The test results of gamma-ray sources experiments indicate that RAIN4HPGe can achieve better energy resolution than the previous DAQ. Compared with V1724, RAIN4HPGe (1 GSPS 12-Bit ADC) has more fine rise-time distribution and better discrimination performance for bulk and very bulk events.

**Table 1. Comparison of RAIN4HPGe and the previous DAQ**

| | Previous DAQ (V1724 [4]) | RAIN4HPGe | |
|---|---|---|---|
| Channel | 8 | 8 | 4 |
| ADC | 100 MSPS 14-Bit | 100 MSPS 14-Bit | 1 GSPS 12-Bit |
| Analog Input (Full-Scale) | 2.25 Vpp / 0.5 Vpp | 1.8 Vpp | 0.8 Vpp |
| Analog Input Bandwidth (MHz) | 40 | ~40 | ~100 |
| ADC ENOB | 11.98 (Manual from CAEN) | 11.49 @ 10 MHz | 8.75 @ 100 MHz |
| Data Transfer Rate (MB/s) | up to 200 (VME 64); 80 (Optical Link) | ~87.5 (Giga-Ethernet) | |
| Energy resolution | (365 ± 15) eV @ 59.54 keV (494 ± 8) eV @ 88.03 keV (558 ± 10)eV @ 122 keV | (362 ± 14) eV @ 59.54 keV (415 ± 5) eV @ 88.03 keV (474 ± 8) eV @ 122 keV | |
| Energy linearity | > 0.999999 | > 0.999999 | |
| Discrimination of bulk and very bulk events | Coarse/Insufficient | | Fine/Sufficient |
| FWHM of rise-time distribution of bulk events | $(2.17\pm0.116)\times10^{-2}$ | | $(1.51\pm0.0179)\times10^{-2}$ |

## 4. Conclusions and future work

Integrated with FPGA, external buffers, readout module, 14-Bit 100 MSPS ADC, and 12-Bit 1 GSPS ADC etc., RAIN4HPGe is adequate for digitalization and readout of pulse signals for CDEX-10. Gamma-ray sources experiments show that RAIN4HPGe can achieve better energy resolution than the previous DAQ. Surface, bulk, and very bulk events from timing amplifier, digitized by 12-Bit 1 GSPS ADC, are also discriminated sufficiently, which is essential for background reduction in dark matter detection.

In the future, RAIN4HPGe will be deployed with CDEX-10 in CJPL and preliminary test will be performed. Discrimination methods of bulk and very bulk events will also be intensively implemented. Tests related to low threshold of the Germanium detector and pulse digitalization (include speed and resolution of the ADC) will also be analysed and carried out.

## Acknowledgements

This work is supported by the National Key Research and Development Program of China (2017YFA0402202).

We would like to thank those who collaborated on the CDEX, and words of deep appreciation go to Professor Yulan Li, Qian Yue and Litao Yang for their invaluable advices, supports and various discussions over the years at the Tsinghua University DEP (Department of Engineering Physics).

We are grateful for the patient help of Yu Xue, Wenping Xue, and Jianfeng Zhang. They are seasoned, full-stack hardware technologists with rich experience of solder and rework in the electronics workshop at DEP.